\documentclass[a4paper,12pt]{article}
\usepackage[utf8x]{inputenc}
\usepackage{amsmath,amssymb,cite,dsfont,mathtools,tensor}
\usepackage{hyperref,verbatim}

\DeclareMathOperator{\Tr}{Tr}
\numberwithin{equation}{section}

\title{{\bf Quantum Entropic Ambiguities: Ethylene}}

\author{A. P. Balachandran$^{a,b}$\footnote{balachandran38@gmail.com},\,, A. R. de Queiroz$^{c}$\footnote{amilcarq@unb.br}\, and S. Vaidya$^d$\footnote{vaidya@cts.iisc.ernet.in} \\
\begin{small}{\it $^a$Department of Physics, Syracuse University, Syracuse, N. Y. 13244-1130, USA}
\end{small}\\
\begin{small}{\it $^b$Institute of Mathematical Sciences, Chennai, India} \end{small} \\
\begin{small}{\it $^c$Instituto de Fisica, Universidade de Brasilia, Caixa Postal 04455, 70919-970, Brasilia, DF, Brazil} \end{small}\\
\begin{small}{\it $^d$Centre for High Energy Physics, Indian Institute of Science, Bangalore, 560012, India}
\end{small}}

\date{\empty}

\begin{document}

\maketitle

\begin{abstract}
In a quantum system, there may be many density matrices associated with a state on an algebra of 
observables. For each density matrix, one can compute its entropy. These are in general different. Therefore 
one reaches the remarkable possibility that there may be many entropies for a given state\footnote{Private 
communication from R. Sorkin}. This ambiguity in entropy can often be traced to a  gauge symmetry emergent 
from the non-trivial topological character of the configuration space of the underlying system. It can also 
happen in finite-dimensional matrix models. In the present work, we discuss this entropy ambiguity and its 
consequences for an ethylene molecule. This is a very simple and well-known system where these notions 
can be put to tests. Of particular interest in this discussion is the fact that the change of the density matrix 
with the corresponding entropy increase drives the system towards the maximally disordered state with 
maximum entropy, where Boltzman's formula applies. Besides its intrinsic conceptual interest, the simplicity 
of this model can serve as an introduction to a similar discussion of systems such as coloured monopoles 
and the breaking of colour symmetry. 
\end{abstract}

\section{Introduction}

Many years ago, Balachandran \emph{et al.} \cite{Balachandran1984,Balachandran1984b} and Nelson 
and Manohar \cite{Nelson:1983bu} discovered that colour symmetry is spontaneously broken in the presence 
of nonabelian GUT monopoles \cite{Balachandran1983}.

Subsequently we discovered that this phenomenon is quite common 
\cite{balachandran1991classical,Balachandran1992c}. It can happen whenever wave functions are sections 
of a twisted bundle over a configuration space $\mathcal{Q}$. If the group $H$ associated to the bundle is 
nonabelian, then it is broken for the same reason that the above mentioned monopole breaks color. 
Examples are diverse and include the following:
\begin{itemize}
  \item[a)] Molecules such as ethylene with nonabelian symmetry $H$ \cite{Balachandran1992c};
  \item[b)] Systems of $N$ identical particles with the representation of a braid or permutation group of dimension $2$ or more \cite{balachandran1991classical};
  \item[c)] QCD \cite{Balachandran1983,Inprep1} and GUT theories \cite{balachandran1983gauge} with their gauge groups as $H$;
  \item[d)] Nonabelian mapping class diffeomorphism groups $H$ of quantum gravity \cite{PhysRevLett.44.1100, PhysRevLett.45.148, Balachandran:2011gj,Inprep2}. 
\end{itemize}

It is often the case that such anomalous quantum breakdown of a classical symmetry is not desirable. With 
that in mind, Balachandran and Queiroz \cite{Balachandran2012} had suggested the use of appropriate 
mixed states which restore the symmetry.

Summarizing, if $H$ is a twisted nonabelian gauge symmetry, then it is anomalous on pure states.
%\footnote{If $\mathds{C}H$ is the group algebra of $H$, and $\mathcal{A}$ a maximal abelian subalgebra of 
%$\mathds{C}H$, then we can quantise so that $\mathcal{A}$ is not spontaneously broken \cite{Balachandran2013a}.}. 
The use of appropriate mixed states removes this anomaly.

The contribution of the present paper is to show that {\it such mixed states necessarily emerge in quantum theory}. We will show the result here for the simple quantum mechanical system of the ethylene molecule $C_2 H_4$. That can be the basis for the future treatment of GUT monopoles and QCD.

The framework best adapted for this analysis is the GNS theory. It formulates quantum theory using unital $C^*$-algebras of observables $\mathcal{A}$ and states $\omega$ on $\mathcal{A}$. The usual Hilbert space formulation emerges from this theory. All results of physical interest can be formulated in the latter, but certain ideas and approaches become less transparent. With this in mind, in Section 2, we explain the aspects of the GNS theory of interest here using the Hilbert space language.

Section 3 recalls the basic quantum theory of ethylene. 

Section 4 deals with the algebra $\mathcal{A}$ of observables of this system. It shows that a ground state wave function induces a {\it mixed} state on $\mathcal{A}$. It is not unique. That reflects the fact that the (convex) set of states is not a simplex \cite{Schroer2009,Balachandran2012f}.

The final section 5 examines what happens on appropriately including the electronic observables in $\mathcal{A}$. Then one can actually prepare the molecule in one of the above mixed states. Further time evolution becomes a stochastic map steadily increasing entropy towards its fixed point which is its maximum \cite{Balachandran2012f}.

\section{The GNS Construction} 

In elementary quantum theory, a general state is identified with a density matrix $\rho$. It is represented in terms of rank $1$ (pure) orthogonal density matrices $\rho_\alpha$:
\begin{align}
  \rho &=\sum_\alpha \lambda_\alpha \rho_\alpha, \qquad \lambda_\alpha > 0, \quad \sum_\alpha \lambda_\alpha = 1, \\
  \rho_\alpha\rho_\beta &=\delta_{\alpha \beta} \rho_\alpha.
\end{align}
These $\rho$ and $\rho_\alpha$, as also the (unital $C^*$-) algebra of observables $\mathcal{A}$, are 
regarded as operators on a Hilbert space $\mathcal{H}$. The expectation value of $a\in\mathcal{A}$ in 
the state specified by $\rho$ is then
\begin{equation}
  \langle a \rangle_\rho = \Tr~\rho a.
\end{equation}
The entropy of $\rho$ is
\begin{equation}
  S(\rho)=-\Tr~\rho\log\rho = -\sum_\alpha \lambda_\alpha\log\lambda_\alpha. 
\end{equation}

We now deduce the GNS language using the density matrix description, first for the rank 1 density matrices and then for general rank ones.

\subsection{Rank $1$ Density Matrix}

In this case,
\begin{align}
	\label{pure-dens-mat-1}
  \rho &=|\psi\rangle\langle \psi |, \qquad \psi\in \mathcal{H}, \nonumber \\
  \langle \psi | \psi\rangle &= 1.
\end{align}

The expectation value of the product $a_1\cdot ... \cdot a_N$ of $N$ observables for this density matrix is
\begin{equation}
	\label{rho-vev-1}
  \Tr~\rho(a_1\cdot ... \cdot a_N) = \langle \psi | a_1\cdot ... \cdot a_N |\psi\rangle.
\end{equation}
Further, for any observable $a$,
\begin{equation}
	\label{positive-inner-prod-1}
  \langle \psi| a^* a |\psi\rangle \geq 0.
\end{equation}

From (\ref{positive-inner-prod-1}) we see that the set of vector states in  $\mathcal{H}$ excited by 
$\mathcal{A}$ from $|\psi\rangle$ is
\begin{equation}
	\label{excited-states-A-1}
  \mathcal{A}|\psi\rangle \subseteq \mathcal{H} = \left\{ |a\rangle_\psi := a|\psi\rangle:~a\in\mathcal{A} \right\}.
\end{equation}
The inner product on these vectors is given by (\ref{positive-inner-prod-1}).

It may happen that certain observables $n$ annihilate $|\psi\rangle$. Such $n$ have zero norm,
\begin{equation}
  \label{null-vectors-1}
  \langle \psi |n^* n|\psi\rangle = 0,
\end{equation}
and generate a left-ideal $\mathcal{N}$ in the algebra, as is readily shown using the Schwarz inequality. 
We must remove this ``Gel'fand ideal'' $\mathcal{N}$ from $\mathcal{A}$ to convert \ref{excited-states-A-1} 
to a Hilbert space. Thus we consider
\begin{equation}
  \left( \mathcal{A}/\mathcal{N}\right)|\psi\rangle = \left\{ |[a]\rangle_\psi:=(a+\mathcal{N})|\psi\rangle:~a\in\mathcal{A} \right\},
\end{equation}
where $\mathcal{N}|\psi\rangle$ denotes the set $\{n|\psi\rangle:~n\in\mathcal{N}\}$. The scalar product for these vectors is given by 
\begin{equation}
	\label{positive-inner-prod-1.2}
  \langle [a]|[b]\rangle_\psi = \langle\psi|a^* b|\psi\rangle.
\end{equation}
Note that $[a]=[a']$, if $a-a'\in\mathcal{N}$. Also by Schwarz inequality $\langle \psi |a^* n | \psi \rangle =0$ 
for $a \in {\cal A}$ so that the RHS of (\ref{positive-inner-prod-1.2}) does not depend on the chosen 
elements $a,b$ from $[a],[b]$.

We can now complete $(\mathcal{A}/\mathcal{N})|\psi\rangle$ using the norm of (\ref{positive-inner-prod-1.2}) to obtain a variant  $\widetilde{\mathcal{H}}_{GNS}$ of the canonical GNS Hilbert space. 
  
On $\widetilde{\mathcal{H}}_{GNS}$, there is also a $*$-representation $\widetilde{\pi}$ of $\mathcal{A}$ (the star becoming the adjoint operator in the representation):
\begin{equation}
  \widetilde{\pi}(b)|[a]\rangle_\psi = |[ba]\rangle_\psi.
\end{equation}

The image of the density matrix $\rho$ in $\widetilde{\mathcal{H}}_{GNS}$ is the vector
\begin{equation}
  |[\mathds{1}]\rangle_\psi.
\end{equation}
(Recall that by assumption $\mathcal{A}$ is unital, that is, $\mathds{1}\in \mathcal{A}$.). The density matrix
\begin{equation}
  \widetilde{\rho}_{GNS}=|[\mathds{1}]\rangle_\psi~_\psi\langle[\mathds{1}]|
\end{equation}
for the algebra $\widetilde{\pi}(\mathcal{A})$ is entirely equivalent to the density matrix $\rho$ of $\mathcal{A}$:
\begin{equation}
  \Tr_{\widetilde{\mathcal{H}}_{GNS}} \widetilde{\rho}_{GNS}\widetilde{\pi}(a) = \Tr_\mathcal{H} \rho a, \quad \textrm{ for all } a\in\mathcal{A}.
\end{equation}

\subsection{Rank $N$ Density Matrix}

In this case,
\begin{align}
  \rho_\alpha&=|\psi_\alpha\rangle\langle \psi_\alpha |, \quad \psi_\alpha\in \mathcal{H}, \nonumber \\
  \langle \psi_\alpha|\psi_\beta\rangle &= \delta_{\alpha\beta},
\end{align}
while (\ref{rho-vev-1}) and (\ref{positive-inner-prod-1}) are replaced by
\begin{align}
  \Tr~\rho\left(a_1\cdot...\cdot a_N \right) &=\sum_\alpha \lambda_\alpha \langle \psi_\alpha|a_1\cdot...\cdot a_N|\psi_\alpha\rangle, \label{rho-vev-2} \\
  \sum_\alpha \lambda_\alpha \langle \psi_\alpha| b^* b|\psi_\alpha\rangle &\geq 0.  \label{positive-inner-prod-2}
\end{align}

The space $\mathcal{A}|\psi\rangle$ is replaced by the direct sum
\begin{equation}
  \bigoplus_\alpha \mathcal{A}|\psi_\alpha\rangle.
\end{equation}

This space inherits the following inner product $(\cdot,\cdot)$ from (\ref{positive-inner-prod-2}):
\begin{equation}
	\label{inner-prod-4}
  \left( a|\psi_\alpha\rangle, b|\psi_\beta\rangle \right) = \sum_\alpha \lambda_\alpha \delta_{\alpha\beta} \langle\psi_\alpha|a^* b |\psi_\alpha \rangle.
\end{equation}
Let us first assume that there are no vectors of zero norm for this inner product. Then the completion of $\bigoplus_\alpha \mathcal{A}|\psi_\alpha\rangle$ using this scalar product gives the full Hilbert space 
$\bigoplus_\alpha \widetilde{\mathcal{H}}_\alpha$, with $\widetilde{\mathcal{H}}_\alpha$ being the completion of $\mathcal{A}|\psi_\alpha\rangle$.  
Thus \emph{$\bigoplus \widetilde{\mathcal{H}}_\alpha$ is an orthogonal direct sum with the weighted inner product (\ref{inner-prod-4})}.

Also while $\mathcal{A}|\psi_\alpha\rangle$ are subspaces $\iota(\widetilde{\mathcal{H}}_\alpha)$ of $\mathcal{H}$ \emph{as vector spaces}, the scalar product (\ref{inner-prod-4}) differs from that inherited by $\iota(\widetilde{\mathcal{H}}_\alpha)$ from $\mathcal{H}$. That is because as subspaces $\iota(\widetilde{\mathcal{H}}_\alpha)$ of $\mathcal{H}$, they would in general have non-trivial intersection for $\alpha\neq \beta$:
\begin{equation}
  \iota(\widetilde{\mathcal{H}}_\alpha) \cap \iota(\widetilde{\mathcal{H}}_\beta) \neq \{ 0 \}
\end{equation}
and hence vectors with non-zero scalar product. Therefore it is not appropriate to regard $\bigoplus_\alpha \widetilde{\mathcal{H}}_\alpha$ as a subspace of $\mathcal{H}$.

Suppose next that there are null vectors. Then we must remove them first from 
$\bigoplus_\alpha \mathcal{A}|\psi_\alpha\rangle$. They come from the $n\in \mathcal{A}$ such that
\begin{equation}
  \langle \psi_\alpha | n^* n|\psi_\alpha\rangle = 0, \quad \textrm{for all } \alpha.
\end{equation}
Such $n$ form the Gel'fand ideal $\mathcal{N}$ of $\mathcal{A}$.

The space $\bigoplus_\alpha \left(\mathcal{A}/\mathcal{N}\right) |\psi_\alpha\rangle$, with
\begin{equation}
\left(\mathcal{A}/\mathcal{N}\right) |\psi_\alpha\rangle := \left\{ (a+\mathcal{N})|\psi_\alpha\rangle \equiv |[a]\rangle_\alpha \right\},
\end{equation}
has scalar product
\begin{equation}
  \langle [a]|[b]\rangle= \sum_\alpha \lambda_\alpha \langle \psi_\alpha|a^* b |\psi_\alpha\rangle.
\end{equation}
On completion it gives a variant $\widetilde{\mathcal{H}}_{GNS}$ of the canonical GNS Hilbert space.

We can write 
\begin{equation}
\mathds{1} = \sum_\alpha |\psi_\alpha \rangle \langle \psi_\alpha|=\sum_\alpha \rho_\alpha
\end{equation}
so that the component of $|[\mathds{1}] \rangle$ in ${\cal A}/{\cal N} |\psi_\alpha \rangle$ is
\begin{equation}
|[\mathds{1}] \rangle_\alpha := \big|[|\psi_\alpha \rangle \langle \psi_\alpha|] \big\rangle_\alpha=|[\rho_\alpha] \rangle_\alpha.
\end{equation}

The Hilbert space $\widetilde{\mathcal{H}}_{GNS}$ carries a $*$-representation $\widetilde{\pi}$ of 
$\mathcal{A}$:
\begin{align}
  \widetilde{\pi}(a)|[b]\rangle &=|[ab]\rangle, \\
  \widetilde{\pi}(b) |[\mathds{1}]\rangle &:= |[b]\rangle.
\end{align}

Since
\begin{equation}
  _{\beta}\langle [c] | \widetilde{\pi}(a) |[b]\rangle_{\alpha}=\delta_{\alpha\beta} \langle 
  \psi_\beta|c^*ab|\psi_\alpha\rangle,
\label{sandwich-1}
\end{equation}
so that each subspace $|[\mathcal{A}]\rangle_{\alpha}$ is invariant under $\widetilde{\pi}$, the density matrix
\begin{equation}
  \widetilde{\rho}_{GNS}=\sum_\alpha |[\mathds{1}]\rangle_{\alpha} ~\sum_\beta ~_{\beta}\langle[\mathds{1}]|
\end{equation}
is entirely equivalent to $\rho$ :
\begin{equation}
  \Tr_{\widetilde{\mathcal{H}}_{GNS}}~\widetilde{\rho}_{GNS}\widetilde{\pi}(a)=\Tr_{\mathcal{H}}~\rho a, \quad \textrm{ for all } a\in\mathcal{A}.
  \label{tr=tr}
\end{equation}
The density matrix $\widetilde{\rho}_{GNS}$ can in fact be rewritten discarding the cross terms in (\ref{tr=tr}) in view of (\ref{sandwich-1}):
\begin{equation}
  \widetilde{\rho}_{GNS}=\sum_\alpha |[\mathds{1}]\rangle_{\alpha} ~_{\alpha}\langle [\mathds{1}]|.
  \label{rho_GNS}
\end{equation}
Since 
\begin{equation}
_{\alpha}\langle [\mathds{1}]| [\mathds{1}]\rangle_{\alpha} = \lambda_\alpha,
\end{equation}
in terms of density matrices 
\begin{equation}
\widetilde{\rho}_{GNS, \alpha} \equiv  \frac{1}{\lambda_\alpha} |[\mathds{1}]\rangle_{\alpha} 
~_{\alpha}\langle [\mathds{1}]|,
\end{equation}
(\ref{rho_GNS}) reads
\begin{equation}
  \widetilde{\rho}_{GNS}=\sum_\alpha \lambda_\alpha ~ \widetilde{\rho}_{GNS, \alpha}.
\end{equation}

\subsection{A Shift in Perspective}

A \emph{state} $\omega$ on a unital $*$-algebra $\mathcal{A}$ is a non-negative linear map from 
$\mathcal{A}$ 
to $\mathds{C}$ normalized to $1$ on $\mathds{1}$ and compatible with the $*$-operation:
\begin{align}
  \omega(a^* a) \geq 0, \quad & \quad \omega(a^*)=\overline{\omega(a)}  \\
  \omega(\mathds{1})=1, \quad & \quad a,\mathds{1}\in\mathcal{A}.
\end{align}
Hence a density matrix $\rho$ defines a state $\omega_\rho$:
\begin{equation}
  \omega_\rho(a)=\Tr~\rho a.
\end{equation}

It can and does happen in quantum theory that many density matrices $\rho_i$ give the same state:
\begin{equation}
  \Tr~\rho_j a = \Tr~\rho_k a, \qquad \rho_j\neq \rho_k \textrm{ for } j\neq k,
\end{equation}
so that
\begin{equation}
  \omega_{\rho_j} = \omega_{\rho_k}.
\end{equation}
But their entropies can be different:
\begin{equation}
  -\Tr~\rho_j \log \rho_j \neq -\Tr~\rho_k \log \rho_k, \quad \textrm{ for } j\neq k.
\end{equation}

We can capture such subtleties more elegantly by not starting with a density matrix, but with a state $\omega$ on $\mathcal{A}$, presented now as a unital $*$-algebra, shifting the focus away from its representation on $\mathcal{H}$ by a density matrix.

In the original GNS approach, one introduces a vector space $\mathcal{A}|\mathds{1}\rangle$ labelled by elements of $\mathcal{A}$:
\begin{equation}
  \mathcal{A}|\mathds{1}\rangle := \left\{ a|\mathds{1}\rangle=|a\rangle:~a\in \mathcal{A} \right\}.
\end{equation}
Then one uses $\omega$ to define inner products:
\begin{equation}
  \left(|b\rangle,|a\rangle \right)=\omega(b^* a).
\end{equation}
If $\omega=\omega_\rho$, then this coincides with $\Tr~\rho\left( b^* a \right)$:
\begin{equation}
  \omega_\rho(b^* a) = \sum_\alpha \lambda_\alpha \langle \psi_\alpha|b^* a|\psi_\alpha\rangle.
\end{equation}

As in the earlier approach, here too there can be null vectors $|n\rangle$:
\begin{equation}
  \langle n | n \rangle = \omega(n^* n) =0.
\end{equation}
The set $\mathcal{N}$ of $n\in\mathcal{A}$ creating the space $\mathcal{N}|\mathds{1}\rangle$ of null vectors is the Gel'fand ideal. We remove them by considering
\begin{equation}
  \left(\mathcal{A}/\mathcal{N} \right) |\mathds{1}\rangle:=\left\{ |[a]\rangle \right\},
\end{equation}
where, as before, by $[a]$ we mean the equivalence class of elements in $\mathcal{A}$ differing by an element of $\mathcal{N}$:
\begin{equation}
  [a]=[b] ~\Leftrightarrow~ a-b\in\mathcal{N}.
\end{equation}

The scalar product in $\left(\mathcal{A}/\mathcal{N}\right)|\mathds{1}\rangle$ is also given by $\omega$:
\begin{equation}
  \langle [b]|[a]\rangle = \omega(b^* a).
\end{equation}
As before, by Schwarz inequality, the RHS does not depend on the choice of the representatives $b$ and 
$a$ from their equivalence classes. The completion of $\left(\mathcal{A}/\mathcal{N}\right)|\mathds{1}\rangle$ gives the standard presentation of the GNS Hilbert space $\mathcal{H}_{GNS}$.

{\it We emphasize that this process of completion can be important, as it is for the ethylene molecule 
treated below in Sections 3 and 4.}

As before, $\mathcal{H}_{GNS}$ carries a representation $\pi$ of $\mathcal{A}$:
\begin{equation}
  \pi(a) |[b]\rangle = |[ba]\rangle.
\end{equation}

We can generate a dense subset of $\mathcal{H}_{GNS}$ by acting with $\pi(a)$'s on 
$|[\mathds{1}]\rangle$. For this reason, $|[\mathds{1}]\rangle$ is called a ``cyclic vector''.

It is an easy check that the density matrix
\begin{equation}
  \hat{\rho}_{GNS} = |[\mathds{1}]\rangle\langle [\mathds{1}] |
\end{equation}
defines exactly the same state as $\omega$ :
\begin{equation}
  \omega(a)=\Tr_{\mathcal{H}_{GNS}}~\hat{\rho}_{GNS} \pi(a).
\end{equation}

We can now quickly recover the decomposition of $\hat{\rho}_{GNS}$ in pure states. Let us reduce $\pi(a)$ into a direct sum of irreducible representation (IRR's),
\begin{equation}
  \pi=\bigoplus_{\alpha,r} \pi^{(\alpha,r)}
\end{equation}
with the corresponding orthogonal direct sum decomposition of $\mathcal{H}_{GNS}$:
\begin{equation}
	\label{Hilbert-decomp-GNS-1}
  \mathcal{H}=\bigoplus_{\alpha,r} \mathcal{H}^{(\alpha,r)}.
\end{equation}
Here if $\alpha\neq \beta$, $\pi^{(\alpha,r)}$ and $\pi^{(\beta,s)}$ are inequivalent:
\begin{equation}
  \pi^{(\alpha,r)}\neq\pi^{(\beta,s)}, \textrm{ if } \alpha\neq \beta,
\end{equation}
while for fixed $\alpha$, $\pi^{(\alpha,r)}$ and $\pi^{(\alpha,s)}$ are equivalent:
\begin{equation}
  \pi^{(\alpha,r)}\simeq \pi^{(\alpha,s)}, \quad r,s = 1,\cdots N_\alpha
\end{equation}
where $N_\alpha$ is the multiplicity of the representation $\alpha$.

If $P^{(\alpha,r)}$ are projectors to $\mathcal{H}^{(\alpha,r)}$, then
\begin{equation}
  |[\mathds{1}]\rangle=\sum_{\alpha,r} P^{(\alpha,r)} |[\mathds{1}]\rangle = \sum_{\alpha,r} |[P^{(\alpha,r)}]\rangle.
\end{equation}
Therefore
\begin{equation}
  \Tr~\hat{\rho}_{GNS}\pi(a)=\sum_{\alpha,r}\Tr~ |[P^{(\alpha,r)}]\rangle\langle[P^{(\alpha,r)}]| \pi(a),
\end{equation}
so that as \emph{states}, we can write
\begin{align}
\hat{\rho}_{GNS}\simeq \rho_{GNS}&:=\sum_{\alpha,r} \sigma^{(\alpha,r)}, \\  
\sigma^{(\alpha,r)} &:=|[P^{(\alpha,r)}]\rangle\langle[P^{(\alpha,r)}]|,
\end{align}
with the corresponding entropy
\begin{equation}
  S(\rho_{GNS})=-\Tr~\rho_{GNS}\log\rho_{GNS}.
\end{equation}

The vectors $|[P^{(\alpha,r)}]\rangle$ may not be of norm 1, which is to be computed using $\omega$. Let
\begin{equation}
\lambda_{\alpha,r} = \Big( |[P^{(\alpha,r)}]\rangle, |[P^{(\alpha,r)}]\rangle \Big) = \omega((P^{(\alpha,r)})^2) = 
 \omega(P^{(\alpha,r)}).
\end{equation}
Then in terms of normalized density matrices 
\begin{equation}
\rho^{(\alpha,r)} = \frac{1}{\lambda_{\alpha,r}} \sigma^{(\alpha,r)}
\end{equation}
we can write
\begin{equation}
\rho_{GNS} = \sum_{\alpha,r} \lambda_{\alpha,r} \rho^{(\alpha,r)}
\end{equation}
and
\begin{equation}
S(\rho_{GNS})= - \sum \lambda_{\alpha,r} \log \lambda_{\alpha,r}
\end{equation}

But if the same IRR $\alpha$ occurs more than once, then the decomposition (\ref{Hilbert-decomp-GNS-1}) 
is not unique. We can replace the subspace $\mathcal{H}^{(\alpha,r)}$ by
\begin{align}
  \mathcal{H}^{(\alpha,r)}(u) &=\mathcal{H}^{(\alpha,s)} u_{sr} \equiv \{\eta^{(\alpha,s)} u_{sr}: \eta^{(\alpha,s)} \in \mathcal{H}^{(\alpha,s)} \},  \label{rotatedspaces} \\
  u^\dagger u &= \mathds{1}. \nonumber
\end{align}
As  $u=e^{i \theta} \mathds{1}$ gives the same orthogonal decomposition for all $\theta$, we have 
$U(N_\alpha)/U(1) \simeq SU(N_\alpha)/\mathbb{Z}_{N_\alpha}$ worth of distinct decompositions in the above.

Here if $\left\{|[\xi_j^{(\alpha,r)}]\rangle\right\}$ is an orthonormal basis for $\mathcal{H}^{(\alpha,r)}$, then an orthonormal basis for $\mathcal{H}^{(\alpha,r)}(u)$ is
\begin{equation}
  \left\{|[\xi_j^{(\alpha,s)}]\rangle~u_{sr}\right\}.
\end{equation}
In quark model language, one says that if $\alpha$ is a colour index, then $r$ is a flavour index.

Since
\begin{equation}
  \pi(a) \mathcal{H}^{(\alpha,r)}(u) \subseteq \mathcal{H}^{(\alpha,r)}(u),
\end{equation}
we can repeat the construction of $\rho_{GNS}$ using projectors $P^{(\alpha,r)}(u)$ on $\mathcal{H}^{(\alpha,r)}(u)$ getting, generically, a new density matrix $\rho_{GNS}(u)$,
\begin{equation}
  \hat{\rho}_{GNS}\simeq \rho_{GNS}(u),
\end{equation}
and the new entropy
\begin{equation}
  S(\rho_{GNS}(u))=-\Tr~\rho_{GNS}(u)\log\rho_{GNS}(u).
  \label{u_entropy}
\end{equation}
Reference \cite{Balachandran2012f} discusses the dependence of the entropy on $u$ in detail. We return to 
it later.

We have now captured the relevant features of states using the original GNS approach as well. 

\subsection{The Emergent ``Gauge'' Symmetry}

For the IRR $\alpha$ with degeneracy $N_\alpha$, the set of unitary transformations $\{u\}$ modulo 
$U(1)$ forms the  group $SU(N_\alpha)/\mathbb{Z}_{N_\alpha} \equiv {\cal G}_\alpha$. It commutes with 
the algebra of observables $\mathcal{A}$. It is therefore a ``gauge'' symmetry. The full gauge symmetry is 
$\times_\alpha {\cal G}_\alpha$. 

Each ${\cal G}_\alpha$ generates an algebra $\mathds{C}{\cal G}_\alpha$, the group algebra of 
${\cal G}_\alpha$, in the commutant $\mathcal{A}'$ of $\mathcal{A}$. We identify $\mathcal{A}'$ with $\bigoplus_\alpha \mathds{C}{\cal G}_\alpha$.

\subsubsection*{Remark}
{\it It is worth pointing out an important subtlety related to the role of $P^{(\alpha,r)}$'s, the projectors to the 
$\mathcal{H}^{(\alpha,r)}$. Purely considerations based on $\mathcal{A}$ allow us to construct only the central 
projectors $P^{(\alpha)} \equiv \sum_r P^{(\alpha,r)}$. Their further "splitting" into different $P^{(\alpha,r)}$'s 
is impossible if one were to make use of only elements $a \in {\cal A}$. To decompose $P^{(\alpha)}$ 
further, we need to enlarge the algebra ${\cal A}$ to a larger algebra ${\cal \bar{A}}$ containing the 
commutant $\mathcal{A}' = \bigoplus_\alpha \mathds{C}{\cal G}_\alpha$. Using $u \in \mathcal{A}'$, we can 
construct the non-central projectors $P^{(\alpha,r)}$ and subsequently subspaces 
$\mathcal{H}^{(\alpha,r)}(u)$ of (\ref{rotatedspaces}). The $u$-dependence of the entropy 
(\ref{u_entropy}) then follows easily.

The above considerations will play an important role in understanding the example of the ethylene 
molecule discussed in the next section.

Alternatively, if we were to restrict ourselves just to the algebra $\mathcal{A}$, we would still be able to deduce 
if the state is mixed or pure by computing the trace of the central projector $P^{(\alpha)}$: if 
${\rm Tr} \, P^{(\alpha)}/{\rm dim} \, \pi^{\alpha} >1$, then the associated state is mixed, or else it is pure. This trace can be computed just from the representation of $\mathcal{A}$ on the Hilbert space.}

The emergence of a gauge symmetry and $\mathcal{A}'$ are among the remarkable insights from the GNS approach and the Tomita-Takesaki theory \cite{haag1992local}.

We will see that the twisted gauge symmetries in the conventional sense are transmuted to the role of $U(N_\alpha)$. Further since only the elements of $\mathcal{A}\cap \mathcal{A}'$, which are contained in the centre of $\mathcal{A}$, are observable, mixed states of the sort in \cite{Balachandran2012,Balachandran2012b} naturally emerge, eliminating the gauge anomalies.

\section{Configuration Space of Ethylene}

Polyatomic molecules can be approximated by rigid shapes in three dimensions at energies much smaller than, say, the dissociation energy of the molecule. Success of molecular spectroscopy, 
which uses the quantum theory of molecular shapes, bears out the validity of this approximation. Molecular shapes possessing some symmetry (subgroups of the rotation group $SO(3)$) are of particular interest, not only because of simplification that group theory offers, but also because a large number 
of interesting molecules possess some symmetry.

Let us briefly describe the configuration space of a molecular shape that has $H \subset SO(3)$ as its symmetry group \cite{Balachandran1992c}. A conveniently chosen configuration $\mathcal{C}_0$ will be denoted as its 
standard configuration. Then all its other configurations can be obtained from $\mathcal{C}_0$ by applying all rotations to it. It is easy to see that its configuration space $\mathcal{Q}$ obtained in this manner is 
\begin{equation}
\mathcal{Q} = SO(3)/H.
\end{equation}
It is multiply-connected if $H$ is discrete. Since ethylene, the molecule we focus on, has a discrete $H$, we assume henceforth that $H$ is discrete.  For concreteness, we think of $\mathcal{Q}$ as the set of {\it right} cosets of $H$ 
in $SO(3)$. 

A convenient way to think of $\mathcal{Q}$ is to recognize that the universal cover of $SO(3)$ is 
$SU(2)$, which allows one to write 
\begin{equation}
\mathcal{Q} = SU(2)/H^*,
\end{equation}
where $H^*$ is the double cover of $H$.  Since $SU(2)$ is simply connected, $\pi_1(\mathcal{Q}) = H^*$. 

The universal cover $\bar{\mathcal{Q}}$ of $\mathcal{Q}$ is $SU(2)$.

The observables for this system are generated by two types of observables:
\begin{enumerate}
\item Functions on $\mathcal{Q}$;

\item Generators of translations on $\mathcal{Q}$ (i.e. generators of physical rotations), or more generally, the group algebra ${\mathbb C}SU(2)$ associated with physical rotations.
\end{enumerate}

To construct the Hilbert space of wave functions for the molecule, it is convenient to start with 
$\bar{\mathcal{Q}} = SU(2)$ and functions on it. These are spanned by components of the rotation matrices $D^j_{\lambda \mu}$, with $j \in \mathbb{Z}^+/2$, and $\lambda, \mu \in \{-j,-j+1, \cdots,j-1,j \}$. The scalar product is 
\begin{equation}
\left( D^{j'}_{\lambda' \mu'}~|~ D^j_{\lambda \mu} \right) = \int_{s \in SU(2)} d \mu(s) 
\bar{D}^{j'}_{\lambda' \mu'}(s) D^j_{\lambda \mu}(s)
\end{equation}
where $d\mu(s)$ is the invariant $SU(2)$ measure. 

The group $\pi_1(\mathcal{Q})$ acts on 
$\bar{\mathcal{Q}}$ by {\it right} multiplication: 
\begin{equation}
g \rightarrow g h^* \quad {\rm for} \quad g \in SU(2), h^* \in H^*. 
\end{equation}
On the other hand, a physical rotation (see item 2 above) induced by $g \in SU(2)$ acts on the left: 
\begin{equation}
	\label{left-action-1}
(U(g) D^j_{\lambda \mu})(s) = D^j_{\lambda \mu}(gs). 
\end{equation}
Since left- and right-multiplications commute, we see that physical rotations (and more generally all physical observables) commute with the action of $H^*$. {\it Thus $H^*$ is the gauge group for this system}. 

Thus molecules provide realistic examples of physical systems with discrete gauge groups.

In quantum theory, wave functions are thus functions on $SU(2)$ that transform by a fixed UIR $\Gamma$ of $H^*$, different $\Gamma$ describing different intrinsic states of the molecule. These are spanned by the matrix elements $D^j_{\lambda \mu}$, with the index $\mu$ now restricted to a subset of $\{ -j \cdots, j \}$. We will henceforth denote the restricted values of $\mu$ by 
$m$. On this basis, the action of $H^*$ is
\begin{equation}
	\label{H-action-1}
D^j_{\lambda m} (s) \rightarrow D^j_{\lambda m}(sh^*) = D^j_{\lambda m'}(s) \Gamma_{m' m}(h^*).
\end{equation}

It is easily seen that functions on $\mathcal{Q}$ are generated by $\sum_m 
\bar{D}^j_{\lambda m} D^{j'}_{\nu m}$ : they are invariant under the $H^*$-action (\ref{H-action-1}).

(The above paragraphs are only meant to indicate how wave functions and functions on $Q$ are constructed in general. The UIR $\Gamma$ may occur more than once for a fixed $j$, and that is not shown here. Also the above presentation is a version with no gauge fixing. The ethylene case, including the problem of gauge fixing, is worked out in full detail below, so that such issues are covered there.)

The Hamiltonian for the system is proportional to the square of the angular momentum, or more realistically, 
\begin{equation}
\mathcal{H} = \sum_{i=1}^3 \frac{J_i^2}{2 I_i}
\label{asymmetrictop}
\end{equation}
where $I_i$ are the three principal moments of inertia and $J_i$ are angular momentum operators which generate rotations on the \emph{left} of $g$ (c.f. (\ref{left-action-1})):
\begin{equation}
  \left( e^{i\theta J_i} D^j_{\lambda m}\right) (g) = D^j_{\lambda m}\left(e^{-i \theta \frac{\tau_i}{2}} g \right), \quad \tau_i={\rm Pauli \,\, matrices}.
\end{equation}
They hence commute with elements of $H^*$.

The Hamiltonian (\ref{asymmetrictop}) is unbounded, and hence defined only on a dense domain in the 
Hilbert space. It is determined by the linear span of $D^j_{\lambda m}$ as we explicitly see below for 
the ethylene.

Our interest is in situations when $H^*$ is non-Abelian. We will restrict our attention to the case of 
ethylene $C_2 H_4$, which is a planar molecule with $H^*$ as the binary dihedral group $D^*_8$:
\begin{equation}
D^*_8 = \{ \pm {\bf 1}_2, \pm i \tau_i :  i=1,2,3 \}.
\label{dihedral}
\end{equation}
The configuration space is $\mathcal{Q} = SU(2)/D^*_8$ with the non-Abelian fundamental group 
$\pi_1 (\mathcal{Q}) = D^*_8$. The domain of the Hamiltonian is fixed to come from the two-dimensional representation of this group. (Here we do not consider the one-dimensional representations of $H^*$.).

This domain can be constructed by starting from the complex linear span in $\sigma$ of the $SU(2)$ rotation matrices
\begin{equation}
D^j_{\sigma \pm m}, \quad \sigma \in \{-(2j+1), \cdots, 2j+1 \}, \quad j \in \mathbb{Z}^+/2 .
\end{equation}
If $h \in D^*_8$, then $h$ acts on these functions by
\begin{equation}
	\label{right-action-1}
D^j_{\sigma m}(g) \rightarrow D^j_{\sigma m}(gh) = \sum_{m' \in \{m,-m\}} D^j_{\sigma m'}(g) D^j_{m'm}(h).
\end{equation}
These are, for fixed $j, \sigma$, two-dimensional representations of $D^*_8$ acting on the indices $\pm m$. 
They are all isomorphic to (\ref{dihedral}). For fixed $j$, there are $(2j+1)/2$ such representations. As explained 
earlier, physical rotations act on the left and commute with the action of $D^*_8$ and are hence 
gauge invariant. 

In the Hamiltonian (\ref{asymmetrictop}), we assume hereafter for simplicity that all $I_i$ are equal to a common $I$. Then the ground state wave functions are obtained from $D^{1/2}_{\sigma, \pm 1/2}$ after gauge fixing (see below).

\subsection{The Algebra $\mathcal{A}$}

It consists of two parts:

\begin{itemize}
\item[a.] Continuous functions $C^0(\mathcal{Q})$ on $\mathcal{Q}$. They are generated by 
\begin{equation}
\sum_{m' \in \{m,-m\}} \bar{D}^j_{\sigma m'} D^{j'}_{\sigma' m'},
\label{basis}
\end{equation}
the bar denoting complex conjugation.

As they are invariant under $D^*_8$ (the two-dimensional matrices are arranged to transform by the same 
$\Gamma$ for all $j,\sigma$), they are functions on $\mathcal{Q}$. Other functions are obtained by taking 
products of linear combinations of (\ref{basis}). 

Note that the first $m'$ can take on $(2j+1)/2$ values, while the second $(2j'+1)/2$ values.

\item[b.] The group algebra $\mathbb{C}SU(2)$.

\end{itemize}

The algebra ${\cal A}$ is generated by $C^0(\mathcal{Q})$ and $\mathbb{C}SU(2)$.

\subsection*{Gauge Fixing}

After gauge fixing, the domain of the Hamiltonian should come from functions of $\mathcal{Q}$. If global gauge fixation is possible, then there will exist a map $\varphi: \mathcal{Q} \rightarrow SU(2)$ so that if $\Pi$ is 
the projection map $SU(2) \rightarrow \mathcal{Q}\simeq SU(2)/D^*_8$, then
\begin{equation}
\Pi \circ \varphi = {\rm identity \,\, map\,\,on \,\,\mathcal{Q}}.
\label{projectortoQ}
\end{equation}
Then we can fix the gauge as follows. Restrict $D^j_{\sigma \lambda}$ to $\varphi(\mathcal{Q})$, then the domain would be spanned by $D^j_{\sigma \lambda}|_{\varphi(\mathcal{Q})}$. But for the case at hand, there is 
no such {\it smooth} global map $\varphi$. Instead, as explained in \cite{Balachandran2012}, we will cover $\mathcal{Q}$ by open sets $\mathcal{Q}_\alpha$ and find maps $\varphi_\alpha: \mathcal{Q}_\alpha \rightarrow SU(2)/D^*_8$ such that $\Pi \circ \varphi_\alpha = {\rm id}|_{\mathcal{Q}_\alpha}$, the identity map on $\mathcal{Q}_{\alpha}$. Then on $\mathcal{Q}_\alpha \bigcap \mathcal{Q}_\beta$ we get transition functions $h_{\alpha \beta} \in D^*_8$: if $q \in \mathcal{Q}_\alpha 
\bigcap \mathcal{Q}_\beta$, then $\varphi_\beta(q) = h_{\beta \alpha} \varphi_\alpha (q)$.

Let us return to (\ref{basis}) and restrict $D^j_{\sigma,\pm m}$ to $\varphi_\alpha(\mathcal{Q}_\alpha)$. In 
this patch, we can mix $\pm m$ by an action of $h \in D^*_8$ as in (\ref{H-action-1}). It changes the 
local section and mixes $\pm m$. Thus there is a quantum internal multiplicity of 2 even though there is no such 
internal multiplicity indicated in ${\cal A}$ \cite{balachandran1991classical}.

But this $h$ {\it cannot} be globally defined, since, $D^*_8$ being non-Abelian, it does not commute in general with $h_{\beta \alpha}$. So in general $h_{\beta\alpha}~h~ h^{-1}_{\beta\alpha} \neq h$ and $h$ cannot be used to implement the $D^*_8$-action in $\mathcal{Q}_\beta$. 

Another way to say this is that $D^*_8$ changes the domain.

Now the puzzle arises: what "symmetry" group preserving the domain of $H$ and commuting with it, mixes 
the indices $m$?

We approach the problem using a state and GNS construction.

\subsection{The GNS construction for $C_2 H_4$} 

The ground state wave functions have $j=1/2$ when all the $I_i$ in (\ref{asymmetrictop}) are equal. As said earlier, we will assume this condition. 

Recall that an automorphism $\gamma$ of ${\cal A}$ acts on any state $\omega$ via duality:
\begin{equation}
\omega(\gamma(a)) := \omega_\gamma (a), \quad \forall a \in {\cal A}.
\end{equation}
In the present case, since the algebra is invariant under the action of the gauge group, it also acts trivially on 
$\omega$. {\it Its apparent lack of gauge invariance 
(if any) when written in terms of wave functions is misleading and should be ignored.}

If $d\mu(g)$ is the invariant volume form on $SU(2)$ normalised to $8\pi^2$,
\begin{equation}
  \int_{SU(2)} d\mu(g) = 8\pi^2,
\end{equation}
then with the scalar product given by
\begin{equation}
	\label{su2-scalar-prod-1}
  \left( D^{j}_{\lambda\mu}~|~D^{j'}_{\lambda'\mu'}\right) = \frac{1}{8\pi^2} \int d\mu(g) \overline{D}^{j}_{\lambda\mu}(g) D^{j'}_{\lambda'\mu'}(g)=\delta_{jj'} \delta_{\lambda\lambda'}\delta_{\mu\mu'},
\end{equation}
we see that
\begin{equation}
  \omega_{++}(\mathds{1})=\left( D^{\frac{1}{2}}_{++}~|~D^{\frac{1}{2}}_{++}\right)=1.
\end{equation}

We can therefore start with the following rank $1$ density matrix $\rho_{++}$ of a ground state for our GNS construction:
\begin{align}
  \omega_{++}(\cdot) &= \Tr~\rho_{++}~(\cdot), \\
  \rho_{+\pm} &= \left|\frac{1}{2};+\pm\right)\left(\frac{1}{2};+\pm\right|, \\
  \left|\frac{1}{2};+\pm\right) &\equiv \left|D_{+\pm}^{\frac{1}{2}}\right).
\end{align}

\subsubsection*{Remark}
{\it 
The Hilbert space with scalar product given by (\ref{su2-scalar-prod-1}) is $\mathcal{H}_{SU(2)}\equiv L^2(SU(2);d\mu(g))$. We denote its kets and bras by $|\cdot)$ and $(\cdot|$.

The vectors and scalar product of the GNS Hilbert space $\mathcal{H}_{GNS}$ are different. The definition of the scalar product in particular involves the state. We denote kets and bras of $\mathcal{H}_{GNS}$ by $|\cdot\rangle$ and $\langle\cdot|$.

Thus in this section, we are working with two Hilbert spaces $\mathcal{H}_{SU(2)}$ and $\mathcal{H}_{GNS}$. Their corresponding ``traces'' are distinguished by subscripts on $\Tr$.}

The algebra $\mathcal{A}$ of observables is invariant under the action of the symmetry group $SU(2)$ acting on the right of $g$ (c.f. (\ref{right-action-1})). Hence for $a\in\mathcal{A}$,
\begin{equation}
  \omega_{++}(a)=\Tr_{\mathcal{H}_{SU(2)}}~\rho_{++} a = \Tr_{\mathcal{H}_{SU(2)}}~\rho_{+-} a \equiv \omega_{+-}(a).
\end{equation}
where $\omega_{+-}$ is the state defined by $\rho_{+-}.$

The states 
\begin{equation}
	\label{mixed-state-1}
  \omega= \lambda ~ \omega_{++} + (1-\lambda) ~ \omega_{+-}, \quad 
  0 \leq \lambda \leq 1
\end{equation}
are thus entirely equivalent to $\omega_{++}$ when restricted to $\mathcal{A}$. It is here that the remark of 
Section 2.4 assumes significance. By enlarging the algebra $\mathcal{A}$ to $\mathcal{\bar{A}}$ which includes the commutant $\mathcal{A}'$, we can distinguish between the $+$ and the $-$ of the second index in $\omega_{+\pm}$.

We have 
\begin{equation}
  \omega(a)= \big[\lambda \left(1/2;++ \right|~a~\left| 1/2;++ \right) +  (1-\lambda) \left(1/2;+- \right|~a~\left| 1/2;+- \right) \big],
\end{equation}
so that $\omega$ seems a \emph{mixed} state invariant under the symmetry or gauge group $SU(2)$. It is represented by the density matrix
\begin{align}
  \rho &= \big[\lambda \left|1/2;++ \right)\left( 1/2;++ \right| + (1-\lambda) \left|1/2;+- \right)\left( 1/2;+- \right| \big]. \label{mixed-state-3}
\end{align}
 
We can now proceed with the GNS construction using the state (\ref{mixed-state-1}). Its next step is the determination of the null space $\mathcal{N}$. If $n\in\mathcal{N}$, then 
\begin{equation}
  \Tr_{\mathcal{H}_{SU(2)}}~\rho (n^* n) = 0
\end{equation}
or
\begin{equation}
  n|1/2;+\pm)=0.
\end{equation}
Hence if
\begin{equation}
  \mathds{P}_+= \left|1/2;++ \right)\left( 1/2;++ \right| + \left|1/2;+- \right)\left( 1/2;+- \right|
\end{equation}
is the projector to the subspace spanned by $|1/2;+\pm)$, then
\begin{equation}
  n=n\left(\mathds{1}-\mathds{P}_+\right).
\end{equation}
Thus the full null space is
\begin{equation}
  \mathcal{N}=\mathcal{A}\left(\mathds{1}-\mathds{P}_+\right).
\end{equation}
Note that $\mathcal{N}$ is a left-ideal as it should be.

It follows that the component of $\mathds{1}\in\mathcal{A}$ in $\mathcal{A}/\mathcal{N}$ is
\begin{equation}
	\label{ethy-id-component-1}
  \left[\mathds{P}_+\right]=\mathds{P}_++\mathcal{N}
\end{equation}
 and the cyclic vector of GNS is
 \begin{equation}
	 \label{ethy-cyclic-vector-1}
   \left| \left[\mathds{P}_+\right]\right\rangle.
 \end{equation}

 \subsection*{Impurity from Gauge Invariance}
 
 We just saw from (\ref{ethy-cyclic-vector-1}) that the cyclic vector or ``ground state'' in the GNS construction 
 is $\left| \left[\mathds{P}_+\right]\right\rangle$. The mean value of $a\in \mathcal{A}$ in this ``ground state'' is
 \begin{align}
   \left\langle \left[\mathds{P}_+\right]\right|a\left| \left[\mathds{P}_+\right]\right\rangle &=
   \Tr_{\mathcal{H}_{SU(2)}}  \rho \mathds{P}_+ a \mathds{P}_+ \nonumber \\ 
&= \big[ \lambda \left(1/2;++ \right|~a~\left| 1/2;++ \right) + (1-\lambda)\left(1/2;+- \right|~a~\left| 1/2;+- \right) \big]. \label{gsvalue}
 \end{align}

 Thus the cyclic vector gives an \emph{impure} state on $\mathcal{A}$ when we take into account the gauge invariance of $\mathcal{A}$.
 
 \emph{This state is a prototype of the mixed states proposed in \cite{Balachandran2012} to restore gauge symmetry.} 

\subsubsection*{Remark}
{\it Although $\mathds{P}_+ \in {\cal A}$, that is not the case for $\rho_{++}$ and $\rho_{+-}$ on the RHS of (\ref{gsvalue}). This expression for precision should be understood as follows. We first replace $\left| 1/2;+\pm \right)$ by sections of the $H^*$-bundle. They will then not be smooth functions on 
$SU(2)/D^*_8$. But the corresponding $[\rho_{++}]$ and $[\rho_{+-}]$ generate vectors contained in ${\cal H}_{GNS}$ which involves the completion in norm of 
${\cal A}/{\cal N}|[\mathds{1}]\rangle$. The state $\omega$ trivially extends to this completion. It is in this sense that (\ref{gsvalue}) is to be understood. Note that since ${\cal A}$ commutes with gauge transformations, the above RHS is independent of the sections used to define $\left| 1/2;+\pm \right)$. For the same reason, use of functions on $SU(2)$ and not sections in $\rho_{+, \pm}$ does not lead to errors.}

 \subsection*{Entropic Ambiguities}
 
 The representation of $\mathcal{A}$ on $\pi(\mathcal{A})|[\mathds{P}_+]\rangle$ is reducible, each subspace 
 \begin{equation}
   \mathcal{H}_{m}\equiv \mathcal{H}_m(\mathds{1})=\pi(\mathcal{A})|[|1/2;+m)(1/2;+m|]\rangle, \quad m=\pm,
 \end{equation}
being invariant under $\pi(\mathcal{A})$. The GNS Hilbert space 
\begin{equation}
  \mathcal{H}_{GNS}=\bigoplus_m \mathcal{H}_m
\end{equation}
is an orthogonal direct sum. Further the representations $\pi$ of $\mathcal{A}$ on $\mathcal{H}_m$ are both isomorphic.

For these reasons, we can make another orthogonal direct sum decomposition
\begin{align}
  \mathcal{H}_{GNS} &= \bigoplus \mathcal{H}_m(u), \\
  \mathcal{H}_m(u) &= \pi(\mathcal{A}) \sum_s u_{ms} \left| \xi_s(\mathds{1}) \right\rangle \equiv 
  \pi(\mathcal{A}) |\xi_m(u)\rangle, \label{rotatedH}\\
  \xi_+(\mathds{1}) &= \frac{1}{\sqrt{\lambda}} \rho_{++}, \quad \xi_- (\mathds{1}) = 
  \frac{1}{\sqrt{1-\lambda}} \rho_{+-}, \quad \langle \xi_m(\mathds{1})| \xi_n (\mathds{1}) \rangle = \delta_{mn}
\end{align}
with $u^\dagger u = 1$, where $s$ is summed over $\pm$. Each $\mathcal{H}_m(u)$ is invariant under $\pi(\mathcal{A})$. Also
\begin{equation}
  \langle \xi_m(u) | \xi_n(u) \rangle = \Tr_{\mathcal{H}_{SU(2)}} \rho \left(\xi_m(u)^* \xi_n(u)\right)= \langle \xi_m(\mathds{1})| \xi_n (\mathds{1}) \rangle = \delta_{mn}.
\end{equation}
as follows from (\ref{rotatedH}).

Now
\begin{equation}
\omega(a) = \Tr_{{\cal H}_{GNS}} \rho(\mathds{1}) a, \quad a \in {\cal A},
\end{equation}
where
\begin{align}
\rho(\mathds{1}) &= \lambda_+(\mathds{1}) | \xi_+ (\mathds{1}) \rangle \langle | \xi_+ (\mathds{1})| + 
\lambda_-(\mathds{1}) | \xi_- (\mathds{1}) \rangle \langle | \xi_- (\mathds{1})|, \\
\lambda_+(\mathds{1}) &= \lambda, \quad \lambda_-(\mathds{1}) = 1-\lambda.
\end{align}

Using the above results, we have also proved elsewhere \cite{Balachandran2012f} that
\begin{align}
  \left\langle [\mathds{P}_+] |~a~|[\mathds{P}_+]\right\rangle &= \sum_m \lambda_m (u) \langle \xi_m(u)|~a~|\xi_m(u)\rangle, \\
  \lambda_m (u) &= \sum_n |u_{mn}|^2 \lambda_n (\mathds{1}). \label{lambda-u-2}
\end{align}
and hence that the density matrix
\begin{equation}
  \rho(u)=\sum_m \lambda_m (u) |\xi_m(u)\rangle\langle \xi_m(u)| := \sum_m \lambda_m(u) \rho_m(u)
\end{equation}
for all $u$ defines the same state on $\mathcal{A}$.

But in general,
\begin{equation}
  \lambda_m(u)\neq \lambda_m(\mathds{1}).
\end{equation}
We can write
\begin{equation}
	\label{GNS-density-matrix-with-u-1}
  \rho(u)=\sum_m \lambda_m(u)\rho_m(u), \qquad \Tr_{\mathcal{H}_{GNS}} \rho_m(u)=1,
\end{equation}
where $\rho_m(u)$ are rank 1 density matrices. Hence the entropy 
\begin{equation}
S(\rho(u)) = -\sum_m \lambda_m(u) \log \lambda_m (u)
\end{equation}
depends on $u$.

When $u$ is changed to, say, $vu$, we get from (\ref{lambda-u-2}), 
\begin{equation}
	\label{stochastic-map-1}
  \lambda_m(vu)=\sum_s |v_{ms}|^2 \lambda_s(u) = T_{rs}(v) \lambda_s(u).
\end{equation}
Accordingly, the entropy is in general changed when $u$ is changed. 

In fact it generally increases as $T(v)$ is a stochastic map \cite{Balachandran2012f}.

The $\rho(\mathds{1})$ for $\lambda=1/2$ (see (\ref{mixed-state-3})) in our example is an exception. It has $\lambda_s(\mathds{1})=1/2$ for both $s$. It is the maximally disordered state where the Boltzmann formula for entropy applies. It is the fixed point of the stochastic map.

\section{Dynamics on $u$: the Electron Cloud}
 
It is clearly interesting to see if we can put dynamics on $u$. If that can be done, the molecular entropy will undergo stochastic maps, tending to increase steadily in time. It will be constant only at the exceptional fixed points. Thus we would have a version of Boltzmann's theorem that entropy in general keeps increasing.

We claim that such dynamics can be induced from that of the electronic cloud. The observations are based on the work of Balachandran and Vaidya \cite{Balachandran:1996jz}.

We assume as usual that all $I_i$ in the Hamiltonian (\ref{asymmetrictop}) are equal, and that the symmetry group $D^*_8$ has the two-dimensional spinorial representation for the molecule.

For definiteness, we assume that the molecule is in the ground state. Then a full eigenstate of the Hamiltonian including the electronic part is
\begin{equation}
  \Psi_r=\sum_{m} D^{\frac{1}{2}}_{rm}\chi^m(\cdots), \qquad r=\pm\frac{1}{2},
\end{equation}
where $\cdots$ denote the electronic variables.

The index $m$ in $\chi^m$ denotes the component of total angular momentum of electrons along the ``third axis'' of the body-fixed frame of the molecule.

The normalization condition on $\Psi_r$ shows that
\begin{align}
  \sum_m \mu_m=1, \\
  \mu_m=\int |\chi^m(\cdots)|^2,
\end{align}
where the integral is over the variables not shown.

If we now perform only molecular observations, we get the mixed state
\begin{equation}
  \sum_{m=\pm} \mu_m \rho_{r,m}.
\end{equation}
 
Comparing with (\ref{GNS-density-matrix-with-u-1}), we see that $\mu_m$ plays the role of $\lambda_m(u)$.

Under a rotation $v$ of the electrons by total angular momentum,
\begin{equation}
  \chi^m(\cdots)\mapsto v_{mr}\chi^r(\cdots)
\end{equation}
  and
  \begin{equation}
    \mu_m \mapsto \mu_m(v)=|v_{mr}|^2\mu_{r}.
  \end{equation}
This matches (\ref{stochastic-map-1}).

Thus by dynamically rotating the electronic angular momentum by electric or magnetic fields, and then restricting observables to the molecular variables, we can steadily evolve the molecular entropy.

\section{Acknowledgments} 
APB was supported by the Institute of Mathematical Sciences, Chennai. ARQ is supported by CNPq under process number 307760/2009-0.

%\bibliography{/home/amilcarq/Dropbox/Main-Amilcar/superbib.bib}
%\bibliographystyle{JHEP}

\providecommand{\href}[2]{#2}\begingroup\raggedright\endgroup

\end{document}